\documentstyle[12pt]{article}

\begin{document}
\input{epsf}

\begin{flushright}
Sept., 1995 \\
TIT/HEP-306/NP\\
\end{flushright}

\begin{center}
{\huge The pion at finite temperature \\}
\hspace*{1.5cm}
{\large Nobuaki Kodama
\footnote{e-mail address: kodama@th.phys.titech.ac.jp} and
Makoto Oka\\
Department of Physics, Tokyo Institute of Technology\\
Meguro, Tokyo 152  Japan}
\end{center}

\begin{abstract}
Properties of the pion are studied at finite temperature with the help of
the PCAC and the QCD sum rule. The pionic mode is treated consistently
with the thermal pions which consist of the ground state at finite
temperature.
The temperature dependence of the
pion decay constant is estimated
for $m_{q}=0$. No consistent solution is found above
the critical temperature $T_{c} =
\sqrt{2} f_{\pi}^{T=0} $. The QCD sum rule shows that the finite quark
mass makes the critical temperature move up to a temperature
where the dilute pion gas
approximation is no more valid.
It also suggests that the pion decay constant
and the continuum threshold decrease by
$\sim 15\%$ and $\sim 10\%$, respectively,
and that the pion mass slightly ($\sim 3\%$) increases
at $T=160MeV$, where the pion gas approximation seems to break down.
\end{abstract}

\newpage

\section{Introduction}
\hspace{5mm}
Behaviors of the hadron properties at finite
temperature have been studied extensively in the context of the
phase transition of QCD. In particular, the temperature
dependencies of hadron
masses, coupling constants and so on
have been calculated in various approaches \cite{method}.
One of them is the QCD sum rule generalized for finite temperature by the
authors of ref.\cite{1}.

The QCD sum rule is based on the operator product expansion (OPE)
of a current correlator and the hadron
duality \cite{SVZ}.
At first \cite{1,matsu}, the Matsubara Green function \cite{matsu1} was
applied for the massless quark
propagator in the OPE of current correlators at finite temperature.
It  was pointed out, however, that the use of the finite temperature Green
function is not fully consistent for the calculation at temperature below
the critical temperature, $T_{c}$ \cite{hatsuda,Dey,Ioffe}. In the
hadronic phase quarks have discrete spectrum due to confinement, so that
the use of the finite temperature
Green function
requires the full range of their
interactions.

On the other hand,
OPE is supposed to achieve a factorization of the soft scale and the
hard scale. The soft scale dynamics is contained in the matrix elements of
local
operators, while the hard scale dynamics is taken into account
in the Wilson
coefficients. Therefore it is natural to include
the temperature dependence only in the matrix elements of operators as far as
$T \sim \Lambda_{QCD} \ll Q$, where $Q$ is the Euclidean momentum of
currents. We conclude that construction of a finite temperature ground
state is essential for the QCD sum rules at finite, but not high,
temperature.

In this paper we investigate the pion at
finite temperature, since the pion is the dominant particle consisting of
the finite temperature ground state.
At temperature, $T<T_{c}$, the particles constructing the
heat bath (finite temperature ground state) are hadrons,
especially the pion, since the heavier states are suppressed by the
Boltzman factor $e^{-m/T}$
at low temperature.

We employ the thermo field dynamics of pions in describing the ground state
\cite{umezawa} and show that the PCAC is valid even
at finite temperature. The
finite temperature ground state $|T\rangle$ gives clear understanding of the
temperature dependence of the state and naturally avoids
a problem of the analyticity at finite
temperature \cite{Ioffe} which was considered to
unjustify the QCD sum rule approach.

The outline of this paper is as follows. In Sect.2, we define
notations and construct the QCD ground state at finite temperature as a
pion gas in the frame work of thermo field dynamics. In Sect.3, we consider
the PCAC in the ground state at finite
temperature constructed in Sect.2. In Sect.4, the temperature dependence of
the pion decay constant is investigated in the chiral limit. There we use
the dilute pion gas approximation to estimate the finite temperature matrix
elements. In Sect.5,
the effects of the finite quark mass to the critical temperature in Sect.4
and other physical parameters of pion are studied by the QCD sum rule
technique. In Sect.6, the conclusion and discussion are given.

\section{QCD ground state at finite temperature}
\hspace{5mm}
We define the creation and annihilation operators of the pion,
$a^{a \dagger}(\mbox{\boldmath$p$})$ and
$a^{a}(\mbox{\boldmath$p$})$, which create or
annihilate a pion with isospin $a$
and the three dimensional momentum $\mbox{\boldmath$p$}$.
These operators satisfy the following relations:
\begin{eqnarray}
 \left[ a^{a} (\mbox{\boldmath $p$}),a^{b \dagger}
 (\mbox{\boldmath $q$}) \right] &=&
 \delta^{3}(\mbox{\boldmath $q$}-\mbox{\boldmath $p$})
 \delta^{ab} \\
 a(\mbox{\boldmath $p$}) | n \rangle &=& \sqrt{n} |n-1 \rangle \nonumber \\
 a^{\dagger} (\mbox{\boldmath $p$}) | n-1 \rangle &=& \sqrt{n} |n \rangle
\end{eqnarray}
where $|n \rangle$ is a state that contains $n$ pions with momentum
$\mbox{\boldmath$p$}$.
The free pion field is defined by
\begin{equation}
\Phi^{a}(x) = \int \frac{d^3 \mbox{\boldmath $k$}}{\sqrt{(2 \pi)^3 2 k_{0}}}
[ a^{a}(\mbox{\boldmath $k$}) e^{-i k x} +
a^{\dagger a}(\mbox{\boldmath $k$}) e^{i k x} ]
\end{equation}
where $k=(k_{0},
\mbox{\boldmath$k$})$ and $k_{0} = \sqrt{\mbox{\boldmath$k$}^{2} +
m_{\pi}^{2}}$.

We express the ground state(vacuum) at finite temperature as $|T
\rangle$, which is regarded as a gas of free pions. Then the
free pion annihilation operator cannot annihilate $| T
\rangle $.
Here we follow the thermo field dynamics formulated by Umezawa et.al.
in ref.\cite{umezawa}. They introduce the Bogoliubov
transformation and define a new operator
$ \tilde{\alpha}^{\dagger}(\mbox{\boldmath$k$}) $
by
\begin{equation}
a(\mbox{\boldmath$k$}) = \cosh \theta \  \alpha(\mbox{\boldmath$k$}) +
\sinh \theta \ \tilde{\alpha}^{\dagger}(\mbox{\boldmath$k$})
\label{eq : bbt}
\end{equation}
Eq.(\ref{eq : bbt}) is canonical transformation.
The mixing angle $\theta = \theta(\mbox{\boldmath$k$},T)$
depends on the momentum $\mbox{\boldmath$k$}$ and the temperature $T$.

The tilde operator annihilates a hole, or
 creates a particle in a state occupied in the finite temperature
 vacuum. In eq.(\ref{eq : bbt}) the annihilation operator in the true
vacuum is divided into an annihilation operator of a particle and
a creation operator of a hole in the
finite temperature vacuum.

Now we define the finite temperature vacuum $| T \rangle$ such that
\begin{equation}
\alpha(\mbox{\boldmath$k$}) | T \rangle = 0, \
\tilde{\alpha}(\mbox{\boldmath$k$}) | T \rangle = 0
\label{eq : tvcon}
\end{equation}
for all $\mbox{\boldmath$k$}$.

We determine $\theta$ by calculating the ground state average of the
particle number density
which must coincide with the Bose-Einstein distribution, i.e.
\begin{eqnarray}
n_{B}(\mbox{\boldmath$k$},T)
= \langle T | a^{\dagger} (\mbox{\boldmath $k$}) a (\mbox{\boldmath
$k$}) |T \rangle /\delta^3 (0)
 = (\sinh \theta)^2
\end{eqnarray}
with $n_{B}(\omega) = 1/(e^{\omega \beta} -1) $ and $\beta=1/T$.
Then
we get
$$
\cosh \theta = (1+n_{B})^{1/2}, \ \sinh \theta = n_{B}^{1/2}.
$$

As we have argued in Introduction, it is regarded
that the quark degree of freedom does not explicitly depend on
temperature in the confined hadron phase.
This assumption means that the temperature effect is only in the
non-perturbative contributions of QCD such as the quark condensate.
This is of course valid only when the temperature is below
the deconfinement phase transition.
We assume furthermore that $|T\rangle$
consists only of pions, since heavier degrees of freedom
are suppressed by $e^{-m \beta}$ at low temperature.

\section{PCAC at finite temperature}
\hspace{5mm}
In order to study properties of the pion at finite temperature, we
consider the following currents,
\begin{eqnarray}
J_{\mu}^{Aa} &=& \overline{q} \gamma_{\mu} \gamma_{5} t^{a} q \\
J_{5}^{a} &=& \overline{q} i \gamma_{5} t^{a} q
\end{eqnarray}
where $t^{a} = \frac{\tau^{a}}{2}$ is the isospin operator.
In the chiral limit the axial-vector current is conserved,
$\partial^{\mu} J_{\mu}^{Aa} = 2 m_{q} \overline{q} i \gamma_{5} t^{a}
q \rightarrow 0$.
In following
we omit the isospin indices $a$
for simplicity.

It is important to note that the
time-ordered correlator is not analytic in the $q^2$ plane at finite
temperature \cite{landau}. Instead, the pionic mode can be studied in
the retarded correlation function,
\begin{eqnarray}
\Pi(q) &=&
\int d^4 x e^{i q x} i \partial_{\mu} \langle T | R[ J_{\mu}^{A} (x),
J_{5}(0) ]| T \rangle \nonumber \\
&=& q_{\mu} \int d^4 x e^{i q x} \langle T | R[ J_{\mu}^{A} (x),
J_{5} (0) ]| T \rangle \nonumber \\
&=& q_{\mu} \int d^4 x e^{i q x} \sum_{n,m} \theta(t) \nonumber \\
&&\left[
e^{-i(p_{n}-p_{m}) x}
\langle T| J_{\mu}^{A} |n \tilde{m} \rangle
\langle n \tilde{m}| J_{5} | T \rangle -
e^{i (p_{n} - p_{m}) x}
\langle T|J_{5}|n \tilde{m}\rangle
\langle n \tilde{m}| J_{\mu}^{A} |T \rangle \right] \nonumber \\
\label{eq : corr1}
\end{eqnarray}
Here we insert the complete set $
1 = \mathop{\sum}_{n,m} |n \tilde{m}\rangle \langle n \tilde{m}| $ as
the intermediate state. $|n \tilde{m}\rangle $ denotes a
$n$-particle $\tilde{m}$-hole state and
$p^{\mu}_{n}$
is the 4-momentum of the
$n$-particle state, while $-p^{\mu}_{m}$ is the 4-momentum of the
$\tilde{m}$-hole state.

The PCAC equation, $\partial^{\mu} J_{\mu}^{A} = 2 m_{q} J_{5}$, and
$\partial^{\mu} \langle T | J_{\mu}^{A} |n \tilde{m} \rangle = - i
(p_{n}-p_{\tilde{m}})^{\mu} \langle
T | J_{\mu}^{A} | n \tilde{m} \rangle $ leads to
$$
\langle T|J_{5}|n \tilde{m}\rangle
\langle n \tilde{m}| J_{\mu}^{A} |T \rangle = -
\langle T|J_{\mu}^{A} |n \tilde{m}\rangle
\langle n \tilde{m}| J_{5} |T \rangle
$$
When we take the rest-frame, i.e. $\mbox{\boldmath$q$} =0 $ and $
q_{0}=\omega$,
the above correlation function is given by
\begin{eqnarray}
\Pi(\omega) &=&
\omega \int_{0}^{\infty} d t \int d \sigma
(e^{i (\omega-\sigma) t}+e^{i ( \omega + \sigma) t}) \nonumber \\
&& \times (2 \pi)^3 \mathop{\sum}_{n,m} \delta(\sigma - (E_{n}-E_{m}))
\delta^3 (\mbox{\boldmath$p$}_{n} - \mbox{\boldmath$p$}_{m})
\langle T|J_{0}^{A}|n \tilde{m} \rangle
\langle n \tilde{m} | J_{5} |T \rangle \nonumber \\
&=& i \int d \sigma \rho(\sigma)
 \frac{2 \omega^2}{( \omega +i \epsilon)^{2} -
\sigma^2}
\label{eq : corr2}
\end{eqnarray}

On the other hand, $\Pi(\omega)$ can be evaluated diredtly, giving
\begin{eqnarray}
\Pi( \omega ) &=&
\int d^4 x e^{i \omega t} i \partial_{\mu}
\langle T | R[ J_{\mu}^{A} (x), J_{5}(0) ]| T \rangle \nonumber \\
&=& \int d^4 x e^{i \omega t} i \delta(t) \langle T | [ J_{0}^{A} (x),
J_{5}(0) ]| T \rangle
\nonumber \\
&=& \langle T | 2 \overline{q} q(0) | T \rangle
\end{eqnarray}
Thus we obtain
\begin{equation}
i \int d \sigma  \rho(\sigma)
 \frac{2 \omega^2}{(\omega +i \epsilon)^{2}- \sigma^{2}} =
2 \langle T | \overline{q} q | T \rangle
\label{eq : pcac1}
\end{equation}

Eq.(\ref{eq : pcac1}) is valid for any
$\omega$, i.e. even at $\omega \rightarrow 0$ and the RHS does not depend
on $\omega$. Then LHS also should not depend on $\omega$.
Therefore, we conclude that
if $ \langle T | \overline{q} q(0) | T \rangle \neq 0 $,
this correlator is
saturated by a massless pole, $\rho(\sigma) = A \delta(\sigma^2)$.
Other states do not contribute to this
correlator at all in the chiral limit.

The spectral function $\rho(\sigma)$ is defined in eq.(\ref{eq :
corr2}),
\begin{eqnarray}
\rho(\sigma) =
(2 \pi)^3 \mathop{\sum}_{m,n} \delta (\sigma-(E_{n}-E_{m}))
\delta^3 (\mbox{\boldmath $p$}_{n}-\mbox{\boldmath $p$}_{m})
\langle T | J_{0}^{A}| n \tilde{m} \rangle
\langle  n \tilde{m} | J_{5} |T \rangle
\label{eq : spec}
\end{eqnarray}
The finite temperature vacuum satisfies $\langle  T | J |T \rangle = 0$
for $J= J_{5} \ $or$\  J_{0}^{A}$ because of parity conservation \footnote{
The finite temperature ground state is connected to the true vacuum by $| T
\rangle = G^{-1}(\theta) |0 \rangle $ with $G(\theta) = exp \left[ \int d^3
\mbox{\boldmath$k$} \theta [ a(\mbox{\boldmath$k$})
\tilde{a}(-\mbox{\boldmath$k$}) - a^{\dagger} (\mbox{\boldmath$k$})
\tilde{a}^{\dagger} (-\mbox{\boldmath$k$})] \right]$.}. Then the lowest energy
contribution to the intermediate states comes from $|n \tilde{m}
\rangle = |1 \tilde{0} \rangle$ and $|0 \tilde{1} \rangle$.
Here we remember that the tilde state is a hole and
has the opposite quantum numbers of a corresponding particle state.
Therefore the
condition $\rho(\sigma) = A \delta(\sigma^2)$ indicates that the two
states
$|1 \tilde{0} \rangle $ and $|0 \tilde{1} \rangle$ represent a massless
particle
and a massless hole, respectively, and are the
only states that contribute in the chiral
limit. These are the pion and its hole at finite
temperature.

One might wonder why there are {\it two } Goldstone modes at finite
temperature. This is caused by the existence of the tilde current
which also satisfies the current conservation law.

At zero temperature a massless pole in the correlator requires existence
of an asymptotic massless field,
$J_{\mu}^{A} \mathop{\rightarrow}_{t \rightarrow
\infty} - \sqrt{2} f_{\pi} \partial_{\mu} \Phi$. Although the finite
temperature vacuum
is not Lorentz invariant \cite{hatsuda}, we expect that the same relation for
$J^{A}_{0}$
is valid in the rest frame at finite
temperature except that the decay constant may depend on temperature, i.e.
$J_{0}^{A} \mathop{\rightarrow}_{t \rightarrow
\infty} - \sqrt{2} f_{\pi}^{T} \partial_{0} \Phi$. Now we find
\begin{eqnarray}
\Pi(\omega) &=&
2 \langle T | \overline{q} q | T \rangle\nonumber \\
 &=&
2i(2 \pi)^3 \left( \langle T | J_{0}^{A}| 1 \tilde{0} \rangle
\langle  1  \tilde{0} | J_{5} |T \rangle +
\langle T | J_{0}^{A}| 0 \tilde{1} \rangle
\langle  0 \tilde{1} | J_{5} |T \rangle \right) \nonumber \\
&=& 2i(2 \pi)^3 \big[
\langle T | - \sqrt{2}  f_{\pi}^{T} \partial_{0} \Phi | 1 \tilde{0} \rangle
\langle  1  \tilde{0} |
\frac{\sqrt{2}  f_{\pi}^{T} (m_{\pi}^{T})^2}{2 m_{q}} \Phi |T \rangle +
\nonumber \\
&&\ \
\langle T | - \sqrt{2}  f_{\pi}^{T} \partial_{0} \Phi | 0 \tilde{1} \rangle
\langle  0 \tilde{1} |
\frac{\sqrt{2}  f_{\pi}^{T} (m_{\pi}^{T})^2}{2 m_{q}} \Phi
|T \rangle \big] \nonumber \\
&=&
- \frac{2 (f_{\pi}^{T})^2 (m_{\pi}^{T})^2}{2 m_{q}}
\left\{ (n_{B}+1) - n_{B} \right\} \nonumber \\
&=&  -2 \frac{(f_{\pi}^{T})^2 (m_{\pi}^{T})^2}{2 m_{q}}
\label{eq : pcacd}
\end{eqnarray}

On the other hand, the retarded correlator of the free pion at finite
temperature is given by
\begin{eqnarray}
&&\langle T | R \left[ \partial_{0}
\Phi(x), \Phi(0) \right] | T \rangle \nonumber \\
&=& \langle T | \int \frac{d^3 \mbox{\boldmath$k$} d^3
\mbox{\boldmath$p$} }{(2 \pi)^3 2 \sqrt{k_{0} p_{0}}} \theta(x_{0})
(-ik_{0})
\nonumber \\
&&( (\sqrt{n_{B}+1} \alpha (\mbox{\boldmath$k$}) e^{- i k x} -
\sqrt{n_{B}} \tilde{\alpha}(\mbox{\boldmath$k$}) e^{i k x})
(\sqrt{n_{B}} \tilde{\alpha}^{\dagger} (\mbox{\boldmath$p$}) +
\sqrt{n_{B}+1} \alpha^{\dagger} (\mbox{\boldmath$p$})) \nonumber \\
&&- (\sqrt{n_{B}+1} \alpha (\mbox{\boldmath$p$}) +
\sqrt{n_{B}} \tilde{\alpha} (\mbox{\boldmath$p$}))
(\sqrt{n_{B}} \tilde{\alpha}^{\dagger} (\mbox{\boldmath$k$}) e^{- i k x} -
\sqrt{n_{B} +1} \alpha^{\dagger}(\mbox{\boldmath$k$}) e^{i k x}) ) |T \rangle
\nonumber \\
&=&\langle T | \int \frac{d^3 \mbox{\boldmath$k$}}{(2 \pi)^3 2 } (-i)
\theta(x_{0}) ((n_{B} +1 ) - n_{B}) \nonumber \\
&&\frac{i}{2 \pi} \int dk_{0} \left\{
\frac{e^{-ik_{0} t} e^{i \mbox{\boldmath$k x$}}}
{k_{0} + i \epsilon -\sqrt{\mbox{\boldmath$k$}^2 + m_{\pi}^2}} +
\frac{e^{-i k_{0} t} e^{- i \mbox{\boldmath$k x$}}}
{k_{0} + i \epsilon + \sqrt{\mbox{\boldmath$k$}^2 + m_{\pi}^2}} \right\}
 |T \rangle
\nonumber \\
&=& \int \frac{dk}{(2 \pi)^4 i} \frac{-i k_{0} e^{- i k x}}
{(k_{0} + i \epsilon)^2 - \mbox{\boldmath$k$}^2 - m_{\pi}^2 }
\label{eq : pprop}
\end{eqnarray}
Clearly the final form does not depend on the temperature and has the same
expression as that at zero temperature. It should be noted that the sum of
the strengths of a particle and a hole state is normalized to the zero
temperature pion propagator.
The difference between eqs. (\ref{eq : pcacd}) and (\ref{eq : pprop})
comes from the temperature dependence of $f_{\pi}$ in the PCAC
relation, $J_{0} = - \sqrt{2} f_{\pi}^{T} \partial_{0} \Phi$.

Finally we obtain the finite temperature version of the
 Gell-Mann-Oaks-Renner relation.
\begin{equation}
\frac{(f_{\pi}^{T})^{2} (m_{\pi}^{T})^{2}}{2 m_{q}} = - \langle T|
\overline{q} q |T \rangle
\end{equation}
This relation is same as that at zero temperature except that LHS means the
sum of a hole and a particle
contributions.

\section{Pion decay constant, $f_{\pi}^{T}$, at finite temperature.}
\hspace{5mm}
We estimate the quark condensate at finite temperature in the dilute pion
gas approximation.
As far as
$T$
is not high, the expectation value of an operator, $O$, at finite
temperature, i.e.
$\langle T | O | T \rangle $,
is given \cite{hatsuda} by
\begin{equation}
\langle T | O | T \rangle  \sim  \langle 0 | O | 0 \rangle  +
\sum^{3}_{a=1} \int d^3 p
\langle \pi^{a}(p)| O |\pi^{a}(p) \rangle n_{B}(\epsilon)
\label{eq : t}
\end{equation}
where
$\epsilon = \sqrt{p^2 + m^2}$, and
$a$
denotes the isospin index. To estimate $\langle \pi|O|\pi \rangle$ we use
the soft-pion theorem.
Since pions which consist of the pion gas are also affected by the heat bath,
the second term in eq (\ref{eq : t}), i.e., the matrix element,
$\langle \pi^{a}(p)| \overline{q}
q |\pi^{a}(p) \rangle$, must
also be temperature dependent. As we have seen in eq.(\ref{eq : pprop}) the
free one-pion
state
$| \pi \rangle =a^{\dagger} |0 \rangle$ does not depend on the temperature.
We conjecture that this effect can be
taken into account by
changing the reduction
formula from $a^{\dagger} \rightarrow \frac{1}{\sqrt{2} f_{\pi}} \partial^{\mu}
J_{\mu}^{A}$ to
$a^{\dagger} \rightarrow \frac{1}{\sqrt{2} f_{\pi}^{T}} \partial^{\mu}
J_{\mu}^{A}$. Then we get
\begin{equation}
\frac{(f_{\pi}^{T})^{2} (m_{\pi}^{T})^{2}}{2 m_{q}}=- \langle \overline{q}
q \rangle_{0} \left[1-\frac{T^2}{8 (f_{\pi}^{T})^{2}} \right].
\label{eq : ftpc}
\end{equation}

\begin{figure}
\epsfbox{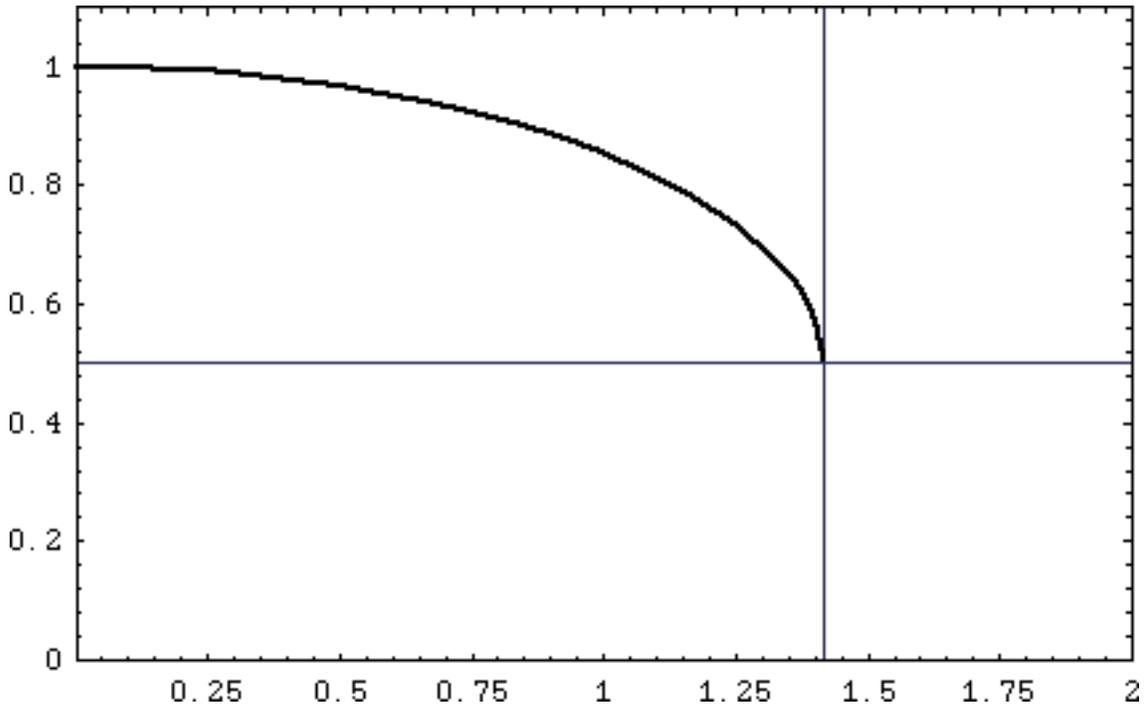}
\caption{$(f_{\pi}^{T}/f_{\pi}^{T=0})^{2}$ v.s. $T/f_{\pi}^{T=0}$}
\label{fig : 0}
\end{figure}

This is easily solved self-consistently under the assumption
$ m_{\pi}^{T}/m_{\pi}^{T=0}=1$ and we find the temperature dependence of
the pion
decay constant,
\begin{equation}
(f_{\pi}^{T})^{2} =\frac{1}{2} \left[ (f_{\pi}^{T=0})^2 +
f_{\pi}^{T=0} \sqrt{(f_{\pi}^{T=0})^2 - \frac{T^2}{2}} \right]
\label{eq : sol}
\end{equation}
The $T$ dependence given by (\ref{eq : sol}) is illustrated in
Fig.\ref{fig : 0}.
It is clear that there exists no solution of eq.(\ref{eq :
ftpc}) above $T > T_{c} = \sqrt{2} f_{\pi}^{T=0}$. $f_{\pi}^{T}$
decreases towards
$f_{\pi}^{T_{c}} = \frac{f_{\pi}^{T=0}}{\sqrt{2}}$ when the temperature
increases from $0$ to
$\sqrt{2} f_{\pi}^{T=0}$.
This singular behavior comes from the consistency condition that the
thermal pions are modified themselves in the finite temperature
ground state. At $T= T_{c}$, the pion gas picture seems to break
down and there is no more consistent solution above $T_{c}$.

\section{QCD sum rules for the pion at finite temperature}
\hspace{5mm}
So far, we have studied the properties of the pion
in the chiral limit, i.e., at $m_{q}=m_{\pi}^{2}=0$.
We study effects of the finite quark mass to the critical temperature and
the other physical parameters of the pion by
the QCD sum rule technique.

\subsection{spectral function}
\hspace{5mm}
We reconsider the spectral function of the
retarded pion correlation function defined by
\begin{equation}
\Pi_{\mu}(q) =
i \int d^4 x e^{i q x} \langle T | R[ J_{\mu}^{A} (x),
J_{5}(0) ]| T \rangle
\end{equation}
Similarly to eq.(\ref{eq : corr1}), we find at the rest frame
\begin{eqnarray}
\Pi_{\mu=0}(\omega) &=&
i \int d^4 x e^{i \omega t} \sum_{n,m} \theta(t) \nonumber \\
&&\times \left[
e^{-i(p_{n}-p_{m}) x}
\langle T| J_{0}^{A} |n \tilde{m} \rangle
\langle n \tilde{m}| J_{5} | T \rangle -
e^{i (p_{n} - p_{m}) x}
\langle T|J_{5}|n \tilde{m}\rangle
\langle n \tilde{m}| J_{0}^{A} |T \rangle \right] \nonumber \\
\end{eqnarray}
Using a relation of the thermo field dynamics \cite{umezawa},
\begin{equation}
\langle T|J_{5}|n \tilde{m}\rangle
\langle n \tilde{m}| J_{0}^{A} |T \rangle = e^{\beta (E_{n} - E_{m})}
\langle T| J_{0}^{A} |m \tilde{n}\rangle
\langle m \tilde{n}| J_{5} |T \rangle
\label{eq : comm}
\end{equation}
we obtain the following expression,
\begin{eqnarray}
\omega \Pi(\omega) &=& \Pi_{\mu=0}(\omega) \nonumber \\
&=&
i  \int_{0}^{\infty} d t \int d \sigma
e^{i (\omega-\sigma) t} (2 \pi)^3 (1 - e^{-\beta \sigma}) \nonumber \\
&&\times \mathop{\sum}_{n,m} \delta(\sigma - (E_{n}-E_{m}))
\delta^3 (\mbox{\boldmath$p$}_{n} - \mbox{\boldmath$p$}_{m})
\langle T|J_{0}^{A}|n \tilde{m} \rangle
\langle n \tilde{m} | J_{5} |T \rangle \nonumber \\
&=& - \int d \sigma \rho_{1}(\sigma)
\frac{1}{\omega -
\sigma + i \epsilon }
\end{eqnarray}
where the spectral function is defined by
\begin{eqnarray}
\rho_{1}(\sigma)
&=& (2 \pi)^3 (1 - e^{-\beta \sigma}) \nonumber \\
&&\times \mathop{\sum}_{n,m} \delta(\sigma - (E_{n}-E_{m}))
\delta^3 (\mbox{\boldmath$p$}_{n} - \mbox{\boldmath$p$}_{m})
\langle T|J_{0}^{A}|n \tilde{m} \rangle
\langle n \tilde{m} | J_{5} |T \rangle
\end{eqnarray}
{}From eq.(\ref{eq : comm}) and the relation
$$
\langle T|J_{5}|n \tilde{m}\rangle
\langle n \tilde{m}| J_{0}^{A} |T \rangle = -
\langle T|J_{0}^{A} |n \tilde{m}\rangle
\langle n \tilde{m}| J_{5} |T \rangle
$$
we obtain the following symmetry relation for the spectral function
$\rho_{1}(\sigma)$,
\begin{equation}
\rho_{1}(\sigma) = \rho_{1}(-\sigma)
\end{equation}
Then we
can rewrite the correlation function,
\begin{eqnarray}
\omega \Pi(\omega^2) &=&
 - \int_{0}^{\infty} d \sigma 2 \omega \rho_{1}(\sigma)
 \frac{1}{(\omega + i \epsilon)^2 - \sigma^2 } \nonumber \\
 &=&  \omega \int_{0}^{\infty} d \sigma \frac{2 \rho_{1}(\sigma)}{
  \sigma^2 - \omega^2 - i \epsilon } \nonumber \\
 &\rightarrow& \omega \int_{0}^{\infty} d \sigma^2 \frac{2 \rho_{1}(\sigma^2)}{
  \sigma^2 - \omega^2 - i \epsilon }
  \label{eq : 26}
\end{eqnarray}
Here the last equality is given for the delta function part in the
spectral function.
Eq.(\ref{eq : 26}) corresponds to eq.(\ref{eq : corr2}) with
$\rho_{1}(\sigma) =
\rho (\sigma) + \rho (-\sigma)$. From eq.(\ref{eq : pcacd}), we expect
that the
strength of the pion
pole in this spectral function, $\rho_{1}(\sigma)$,
is given by $\frac{(f_{\pi}^{T})^2 (m_{\pi}^{T})^2}{2 m_{q}}$.
Then we may assume the spectral function, $\rho_{1}(\sigma^2)$, in the
following form for a finite quark mass,
\begin{equation}
2 \rho_{1}(\sigma^2) = \frac{2 (f_{\pi}^{T})^2
(m_{\pi}^{T})^2}{2 m_{q}} \delta(\sigma^2 - (m_{\pi}^{T})^2) +
Perturbation \cdot \theta(\sigma^2 - s_{0})
\end{equation}
The $Perturbation$ term is estimated by the bare-loop calculation.
This form is same as that at the zero temperature
except that the pion pole consists of
a particle state and a hole state.

{}From our derivation of the spectral function, it is clear that
the correlation function,
$\Pi(\omega^2)$, has an analytic property in the $\omega^2$ plane.
Therefore we use the QCD sum rule techniques for this correlation function in
the usual manner.

Here we write down the imaginary part of correlation function,
$\Pi(\omega^2)$.
\begin{equation}
\frac{1}{\pi} \mbox{Im} \Pi(s) = \frac{2 (f_{\pi}^{T})^2
(m_{\pi}^{T})^2}{2 m_{q}} \delta(s - (m_{\pi}^{T})^2) +
Perturbation \cdot \theta(s - s_{0})
\label{eq : phen}
\end{equation}

\subsection{The construction of the sum rule.}
\hspace{5mm}
Next we construct the theoretical side of the QCD sum rule.
The scalar part of this correlator can be
easily calculated using the Ward
identity and the pseudo-scalar correlator \cite{bible}.
\begin{eqnarray}
\omega \Pi(\omega^2)&=& \nonumber \\
 - \omega &\Big[& (2 m_{q}) \Big\{ \frac{3}{8 \pi^2}
log(-\omega^2/\mu^2) \big\{ 1+ \frac{\alpha_{s}}{\pi}
\big( \frac{17}{3} - log(-\omega^2/\mu^2)
\big)  \big\} +
\frac{\alpha_{s}}{8 \pi \omega^4} G_{\mu \nu} G_{\mu \nu} \nonumber \\
&&- \frac{4 \pi \alpha_{s}}{\omega^6} \big\{ \overline{d}
\gamma_{5} \sigma_{\mu \nu} t^{a} u \overline{u} \gamma_{5} \sigma_{\mu
\nu} t^{a} d  - \frac{1}{3}  (\overline{u} \gamma_{\mu}
t^{a} u + \overline{d} \gamma_{\mu}
t^{a} d) \sum \overline{q} \gamma_{\mu}
t^{a} q \big\} \Big\} \nonumber \\
&&- \frac{2}{\omega^2}  \overline{q} q
\Big].
\end{eqnarray}

At finite temperature the OPE side contains
operators carrying spin \cite{hatsuda}. We do not consider
operators which have Lorentz indices except for the
operators of dimension 3,
since those operators are proportional to higher powers of temperature
and also are multiplied by the small quark mass.

The only possible dimension 3 non-scalar operator is $ \mbox{\it
S.T.}(\overline{d} \gamma_{\mu} \gamma_{\nu} d + \overline{u}
\gamma_{\nu} \gamma_{\mu} u)$, where $ \mbox{\it
S.T.}$ means the symmetric and traceless tensor. This happens to be
zero in the isospin $SU(2)$ symmetry limit, and therefore no spin
operator contributes in this case.

Same as the phenomenological side,
we only consider the same theoretical side with the zero temperature sum
rule except for the temperature dependent condensate.

As is mentioned previously,
as far as $T$ is not so high and the
thermal pion gas is dilute, $\langle T | O(\mu^2) | T \rangle $ is approximated
as
$$
\langle T | O(\mu^2) | T \rangle \simeq \langle 0 | O(\mu^2) | 0 \rangle +
\sum^{3}_{a=1} \int d^3 p
\langle \pi^{a}(p) | O(\mu) | \pi^{a}(p) \rangle n_{B}(\epsilon/T),
$$

We estimate the temperature dependence of the condensate in the soft pion
limit\cite{hatsuda}. We
write down only the results.
$$
\langle \overline{q} q \rangle_{T}= \langle \overline{q} q \rangle_{0}
(1-\frac{T^2}{8 (f_{\pi}^{T})^2} B_{1}(\frac{m_{\pi}^{T}}{T}))
$$
$$
\langle G \cdot G \rangle_{T} = \langle G \cdot G \rangle_{0}
$$
$$
\langle \overline{d}
\gamma_{5} \sigma_{\mu \nu} t^{a} u \overline{u} \gamma_{5} \sigma_{\mu
\nu} t^{a} d \rangle_{T} - \frac{1}{3} \langle (\overline{u} \gamma_{\mu}
t^{a} u + \overline{d} \gamma_{\mu}
t^{a} d) \sum \overline{q} \gamma_{\mu}
t^{a} q \rangle_{T} = -\frac{28}{27} \langle \overline{q}q
\rangle_{0}^{2} (1- \frac{3 T^2}{14 (f_{\pi}^{T})^2}
B_{1}(\frac{m_{\pi}^{T}}{T}))
$$
Here the function $B_{1}(z)$ is defined by
$$
B_{1}(z) = \frac{6}{\pi^2} \int_{z}^{\infty} dy \sqrt{y^2-z^2} (e^y -1)^{-1}.
$$

Equating the theoretical side and the phenomenological side of
the correlation function with the help of
the dispersion relation, and making the
Borel transformation, we obtain the sum rule for the pion at finite
temperature:
\begin{eqnarray}
 \frac{2 (f_{\pi}^{T})^{2} (m^{T}_{\pi})^{2}}{2 m_{q}}
 \frac{\exp^{-(m_{\pi}^{T})^{2}/M^{2}}}{M^2} &=&
 2 m_{q} [ \frac{3}{8 \pi^2}
 (1+ (\frac{17}{3} + 2 \gamma_{E}) \frac{\alpha_{s}}{\pi}) (1-\exp
 (-s_{0}^{T}/M^2)) \nonumber \\
&&- \frac{\alpha_{s}}{8 \pi M^4} \langle GG \rangle_{0}
- \frac{56 \pi \alpha_{s}}{27 M^6} \langle \overline{q} q
\rangle^{2}_{0} (1- \frac{3 T^2}{14 (f_{\pi}^{T})^2}
B_{1}(\frac{m_{\pi}^{T}}{T}) )]
\nonumber \\
&&- \frac{2}{M^2} \langle \overline{q} q \rangle_{0}
(1-\frac{T^2}{8 (f_{\pi}^{T})^2} B_{1}(\frac{m_{\pi}^{T}}{T})).
\label{sum : sum}
\end{eqnarray}

\subsection{ The estimation of temperature dependence.}
\hspace{5mm}
 In its early stages, the QCD sum rule for the pion was known to have a
 difficulty mainly because the coupling of the pion pole is not dominant.
 For the pseudo-scalar correlator, the excited pion states and the
 instanton effects become important, while for the axial-vector correlator,
 the axial-vector meson $a_{1}$ will disturb the sum rule.

 These difficulties, however,
can be circumvented by adapting the pseudoscalor-axialvector
off-diagonal correlator since it is
almost saturated by the pion. We, however,
should not take a logarithmic derivative of the
lowest moment sum rule to eliminate the coupling strength
since the logarithmic derivative strongly enhances the contribution
from the higher resonances. Therefore we evaluate the values of physical
parameters by fitting
the LHS and RHS of the sum rule.

The fitting analysis of the sum rule is performed
in an interval of $M$,  $\Omega
(M_{min} < M < M_{max})$, the lower and upper limits of which are
determined in the following argument.

According to the Ward identity,
the deviation from the chiral limit is given by the pseudo-scalar
correlator. The convergence of the OPE of the off-diagonal correlator
depends on one of the pseudo-scalar
correlator. Then we set the lower limit to ensure that the power correction of
the pseudo-scalar correlator is smaller than $5\%$
of the perturbative contribution. This condition is also expected to suppress
the instanton contribution in the pseudo-scalar correlator.
The upper limit is set to suppress the continuum contribution
to less than $20 \%$ of the total in the pseudo-scalar correlator.

Because $\Omega$ generally depends on temperature, we
determine $\Omega$ at each temperature.
At zero temperature the above conditions give at $s_{0} \sim 2 GeV^2$,
$$\Omega(0.99GeV < M < 1.18GeV)$$

We determine $m_{\pi}^{T}$ and $\lambda(T)=
2 (f_{\pi}^{T})^{2} (m_{\pi}^{T})^{2}$ by matching the RHS,
$R(s_{0}^{T},T,M)$, and the LHS,
$L(m_{\pi}^{T},\lambda(T),M)$ of the sum rule in the
interval $\Omega$.
Let us rewrite the sum rule in the form
$$
\lambda(T) \equiv F(m_{\pi}^{T},s_{0}^{T},T,M)=
(m_{u} + m_{d}) R(s_{0}^{T},T,M) M^2
\exp((m_{\pi}^{T})^2/M^2)
$$
The coupling parameter $\lambda(T)$
should not depend on $M$.
Therefore we choose $m_{\pi}^{T}$ and $s_{0}^{T}$
so that $F$ is least dependent on $M$ in $\Omega$ and
$\lambda(T)$ is determined as the mean value
$\overline{F}(m_{\pi}^{T},s_{0}^{T},T)$ of the function
$F(m_{\pi}^{T},s_{0}^{T},T,M)$ in $\Omega$.
 Then we define
$$
\delta=Max[|F(m_{\pi}^{T},s_{0}^{T},M,T)-
\overline{F}(m_{\pi}^{T},s_{0}^{T},T)|/\overline{F}(m_{\pi},s_{0}^{T},T)]
$$
which represents the variation of $F(m_{\pi}^{T},s_{0}^{T},T,M)$ in $\Omega$.
At a given temperature we search the best fit parameters which minimize
$\delta$.

Because
the condensate depends on the $m_{\pi}^{T}$ and $f_{\pi}^{T}$ at finite
temperature, we have to determine $m_{\pi}^{T}$ and $f_{\pi}^{T}$
self-consistently, i.e., until the difference between the
input $m_{\pi}^{T}$,
$f_{\pi}^{T}$ and the output $m_{\pi}^{T}$,
$f_{\pi}^{T}$ becomes less than $1\%$.

The results are shown in Figures 2-5. We find the similar behavior of
$f_{\pi}^{T}$ as that in the chiral limit. The consistent solution of
$f_{\pi}^{T}$ disappears above the critical temperature $T_{c} \simeq
180MeV$. At $T_{c}$, $m_{\pi} = 146MeV$,
$s_{0} = 1.93$ and $f_{\pi} = 70.9MeV$. The critical value of the pion
decay constant is almost the same as that in the chiral limit.
It is, however, generally suggested that the pions
can no more dominate the finite temperature ground state
at
$T>160MeV$ \cite{gerber}. Therefore we trust our results only
at $T<160MeV$. In this region the pion mass is
almost stable, the decay constant decreases by$\sim 15\%$ and the
continuum threshold
decreases by $\sim 10\%$.

\begin{figure}
\epsfbox{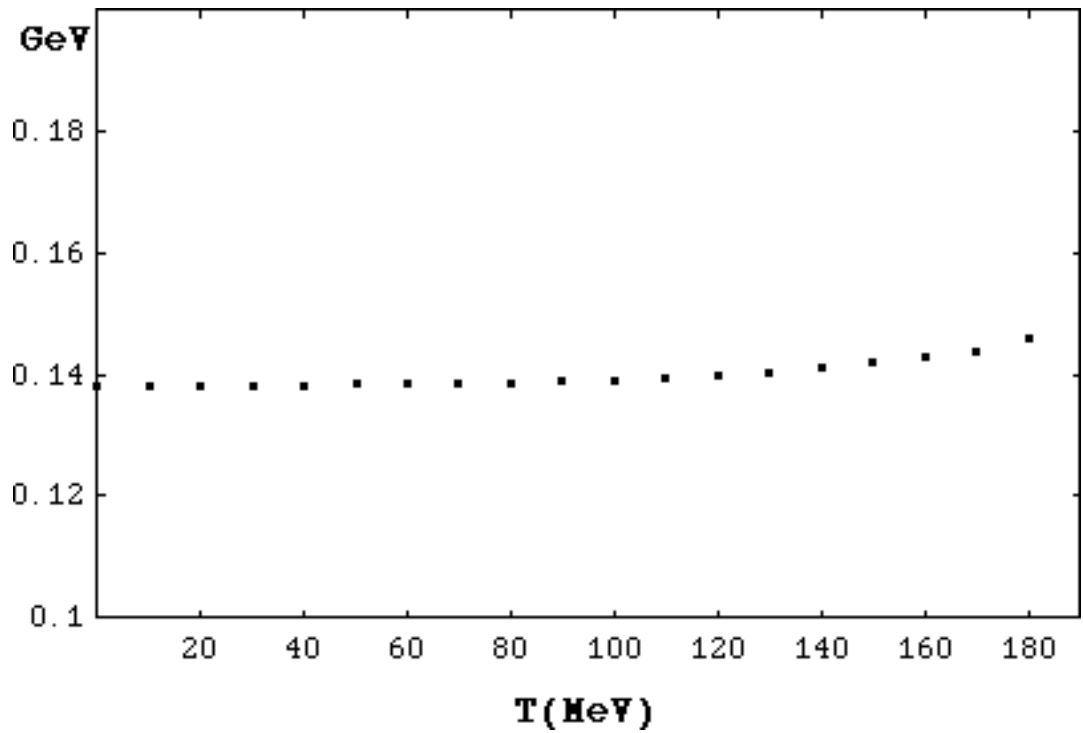}
\caption{The temperature dependence of the pion mass.}
\end{figure}
\begin{figure}
\epsfbox{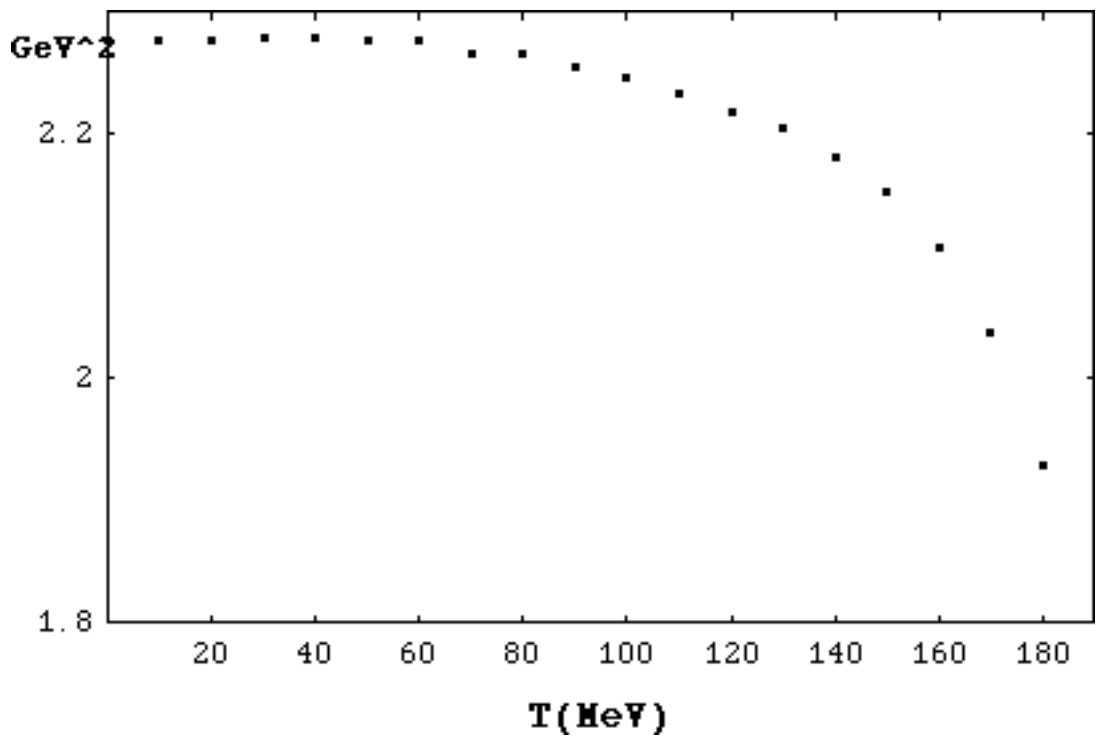}
\caption{The temperature dependence of the continuum threshold $s_{0}$. }
\end{figure}
\begin{figure}
\epsfbox{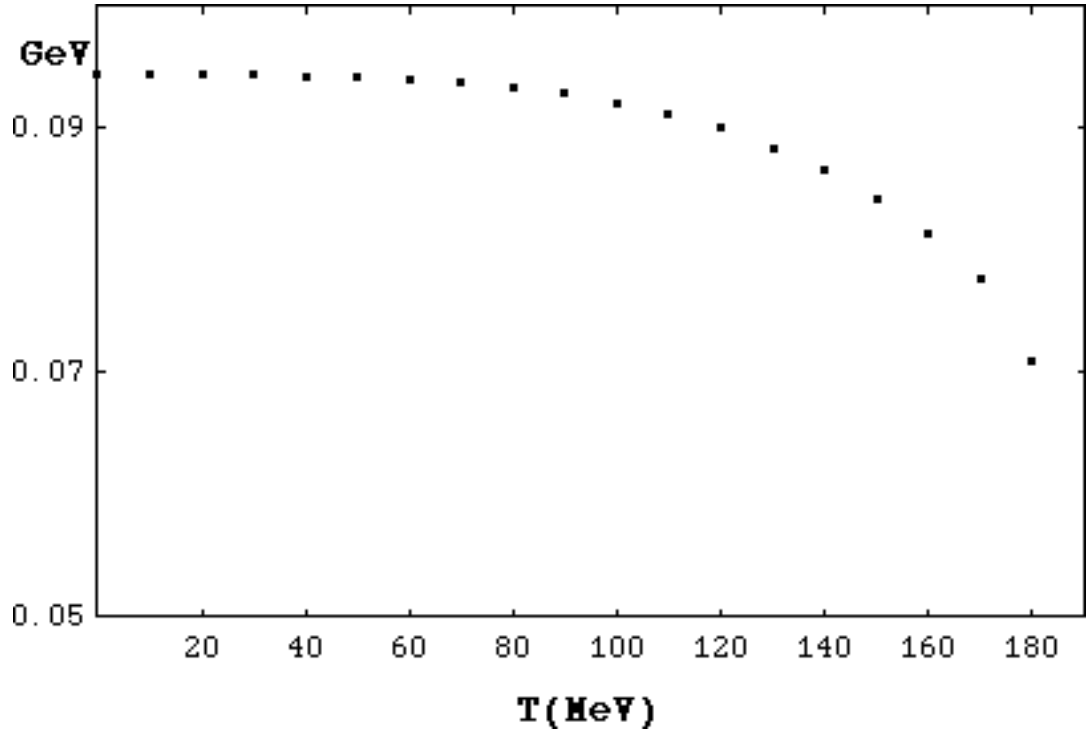}
\caption{The temperature dependence of pion decay constant $f_{\pi}^{T}$.}
\end{figure}
\begin{figure}
\epsfbox{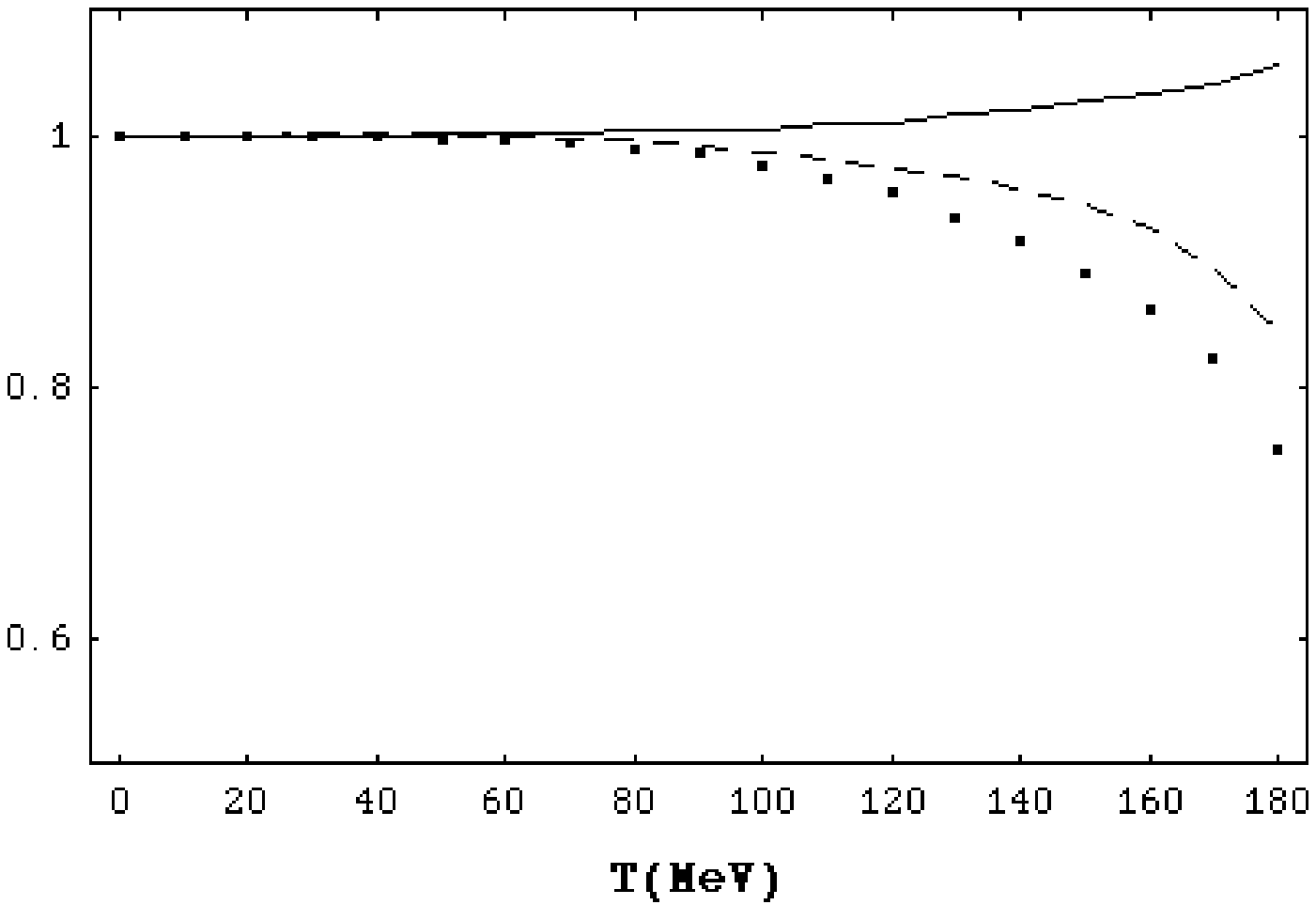}
\caption{The ratios $m_{\pi}^{T}/m_{\pi}^{T=0}$(solid line),
$s_{0}^{T}/s_{0}^{T=0}$ (dashed) and the $f_{\pi}^{T}/f_{\pi}^{T=0}$ (dotted).}
\end{figure}

In this analysis the quark mass and quark condensate are determined
by the pion sum rules  at zero temperature and the
4-quark condensate is evaluated in the rho meson sum rule.
These values are $m_{u} + m_{d} =15MeV$,
$\langle \overline{q} q \rangle_{0} = -
(220MeV)^3$, $\langle \overline{q} q \rangle^2_{0} = 7.3 \cdot 10^{-4}
GeV^6$, and the gluon condensate $\langle
\frac{\alpha_{s}}{\pi} G \cdot G \rangle_{0} = (360MeV)^4$.

\section{Conclusion and Discussion}
\hspace{5mm}
The QCD sum rule is applied to the analysis of the pion properties at
finite temperature.
We find that the pion decay constant decreases, while
the pion mass increases as the temperature increases.
These results qualitatively
coincide with those in other approaches, such as the NJL model \cite{NJL}.
The increase of the pion mass is very small, i.e., $\sim 3\%$
at $T=160MeV$, which is again consistent with the result
in the chiral perturbation theory \cite{pert}.

It is found that the consistency of the pion properties with those of the
thermal pions suggests a critical behavior of the pion decay constant.
In the chiral limit the critical temperature, $T_{c}$ is $ \sqrt{2}
f_{\pi}^{T=0} \sim 133MeV$ for $f_{\pi}^{T=0} =94 MeV$. Above
$T_{c}$, there exists no consistent solution of $f_{\pi}^{T}$.
The higher power corrections of $T$ for
$\langle \overline{q} q \rangle_{T}$ can be expressed
in the chiral limit as follows \cite{gerber}.
\begin{equation}
\langle \overline{q} q \rangle_{T}=\langle \overline{q} q \rangle_{0} (
1 - \frac{T^2}{8 f_{\pi}^2} - \frac{T^4}{ 384 f_{\pi}^4} - \frac{T^6}{288
f_{\pi}^6} ln \frac{\Lambda}{T} +...)
\end{equation}
One sees that they tend to accelerate the breakdown.
Furthermore the contribution of the higher resonances is expected to
suppress the quark condensate. Therefore these corrections to
our analysis are expected to enhance the critical behavior of the pion decay
constant.

Our results of the ''phase transition'' seems to be the first order phase
transition. The QCD phase transition with two massless quarks is expected
to be the second order \cite{sigama}. We conjecture that the
discrepancy comes from the breaking of hadronic description of the finite
temperature vacuum. At the temperature near $T_{c} (= \sqrt{2}
f_{\pi}^{T=0})$, the quark degrees of freedom
are no more frozen, and the self-energy of quarks will
break the PCAC relation. Such an effect might make the
''phase transition''  smooth.

The QCD sum rule with a finite quark mass indicates the same
critical behavior at $T=180MeV$. The difference of the critical
temperature from the chiral limit
can be understood from the consideration that the small
difference of the quark mass leads to a large (compared with the
quark mass) scale
difference in the hadronic world, such as the pion mass.
This, however, may not
justify the hadronic description of the finite temperature
vacuum at $T=180MeV$, since the higher resonance contribution is expected to
dominate the number density at $T>160MeV$.

It is worthwhile to mention about the relation between $s_{0}^{T}$ and
$\langle \overline{q} q \rangle_{T}$. The pion mass is approximately
calculated by taking a logarithmic derivative of eq.(\ref{sum : sum}). If
we omit a few negligible terms and take $M \sim 1GeV$ then we get
$$
(m_{\pi}^{T})^{2} = \frac{2 m_{q} \frac{3}{8 \pi^2} \left[ 1- (1+s_{0}^{T})
e^{-s_{0}^{T}} \right]}{-2 \langle \overline{q} q \rangle_{T}}
$$
When the pion mass increases at finite temperature this equation gives the
following constraint:
$$
-2 \langle \overline{q} q \rangle_{T} (m_{\pi}^{T=0})^{2} <
2 m_{q} \frac{3}{8 \pi^2} \left[ 1- (1+s_{0}^{T})
e^{-s_{0}^{T}} \right]
$$

If we regard the continuum
threshold $s_{0}$ as an order parameter for
the deconfinement phase transition \cite{deco}, then the above constraint
indicates that the critical temperature $T_{d}$, defined by
$s_{0}^{T=T_{d}}=0$,
cannot be lower than the chiral transition temperature $T_{ch}$ with
$\langle \overline{q} q \rangle_{T=T_{ch}}=0$. If the pion
mass does not depend on the temperature, these two temperatures are equal
in our estimation.

\end{document}